\documentclass[%
 reprint,
floatfix,
superscriptaddress,
showpacs,preprintnumbers,
 amsmath,amssymb,
 aps,
pra,
]{revtex4-2}
\DeclareUnicodeCharacter{2212}{-}
\usepackage{xcolor}
\usepackage{times}
\usepackage{amssymb}
\usepackage{dsfont}
\usepackage{graphicx}% Include figure files
\usepackage{dcolumn}% Align table columns on decimal point
\usepackage{bm}% bold math
\usepackage{epstopdf}
\usepackage{lineno}
\usepackage[colorlinks]{hyperref}% add hypertext capabilities
\hypersetup{colorlinks = true,
linkcolor = blue,
filecolor = blue,
urlcolor = blue,
citecolor = blue}
\newcommand*\chem[1]{\ensuremath{\mathrm{#1}}}

\begin{document}

\title{\textbf{Charge density wave solutions of the Hubbard model in the composite operator formalism}}
\author{Anurag Banerjee}
\affiliation{Universit\'{e} Paris-Saclay, Institut de Physique Th\'eorique, CEA, CNRS, F-91191 Gif-sur-Yvette, France}

\author{Emile Pangburn}
\affiliation{Universit\'{e} Paris-Saclay, Institut de Physique Th\'eorique, CEA, CNRS, F-91191 Gif-sur-Yvette, France}

\author{Chiranjit Mahato}
\affiliation{Indian Institute of Science Education and Research Kolkata, Mohanpur, India-741246}

\author{Amit Ghosal}
\affiliation{Indian Institute of Science Education and Research Kolkata, Mohanpur, India-741246}

\author{Catherine P\'epin}
\affiliation{Universit\'{e} Paris-Saclay, Institut de Physique Th\'eorique, CEA, CNRS, F-91191 Gif-sur-Yvette, France}

\begin{abstract}
We investigate the charge density wave phase in the strongly correlated Hubbard model without any other broken symmetry phase. Starting from the atomic Hamiltonian with no hopping, we generate quasiparticle operators corresponding to holons and doublons in the strongly correlated limit of the repulsive Hubbard model. We develop a real space composite operator formalism using the equation of motion technique to include the intersite hopping perturbatively. Our fully self-consistent calculation stabilizes multiple unidirectional translation symmetry broken states within the doping range $\delta=0.07$ to $0.2$. The charge-ordered states become increasingly unfavorable with hole-doping. The unidirectional density waves manifest as periodic modulations of half-filled Mott regions separated by hole-rich regions. Notably, density wave solutions with periods of $3$ to $8$ lattice spacing remain energetically higher than those with larger periods. Quenched disorder on the charge-ordered states induces the merging of the Mott regions and, consequently, forms short ranged charge modulations. The density of states shows signatures of strongly correlated Mott regions, potentially relevant to the physics of underdoped cuprates.
\end{abstract}
\maketitle
\section{Introduction}
Density wave orderings are common in quantum materials such as cuprates~\cite{frano2020charge,VojtaCDWRev}, transition metal dichalcogenides~\cite{YanSTM}, Kagome materials~\cite{Jiang2021,KagomePRB}, and twisted layered systems~\cite{LiangTBG}. Particularly, in hole-doped cuprates, signatures of charge ordering are now well established  ~\cite{Parker2010,Dopig_CupratesCDW,Abbamonte2005,frano2020charge,CDWUbiquitous,RIXSReview,NeutronLaStripes,Wen2019,Comin2015,Cai2016}. Studies at a around $(1/8)$ hole doping provide evidence of spatially modulated Mott (half-filled) regions~\cite{Abbamonte2005}. Across the different families the reported value of the ordering wavelength varies between $3$ to $6$ lattice spacings, $a_0$ over the entire doping range~\cite{frano2020charge,VojtaCDWRev}. Generally, as doping increases, the ordering wavelength also increases~\cite{frano2020charge} (except in La-based compounds), and superconducting correlations dominate. Recently, phase-sensitive measurements on various compounds indicate that bond-centered charge modulation best fits with a dominant d-form factor~\cite{Comin2015}, although other CDW form factors are also reported~\cite{McMahonSymm}. Interestingly, multiple evidences suggest that CDW ordering exists independently from spin ordering, either at high temperatures~\cite{CDW_SDW_La} or under strong magnetic fields~\cite{JangIdealCDW,Wu2011}.

The Hubbard model is the fundamental framework for studying strongly correlated systems and is widely regarded as the primary model for understanding cuprate superconductors.~\cite{AnnualRev_CompHubb,HubbAnnRev_Kiv}. As no exact solutions exist for the doped Hubbard model in two dimensions~\cite{AnnualRev_CompHubb,HubbAnnRev_Kiv}, approximate numerical solutions often vary depending on accessible system size, temperature, and approximation schemes, making it challenging to predict ground state properties at finite dopings. To address such uncertainties, multimethod consensus~\cite{PRX_Hubbard,Stripes2017} studies are being conducted to clarify the key features of the doped Hubbard model solutions.

Stripes, characterized by periodic modulations of antiferromagnetic (AFM) regions where holes accumulate at AFM domain walls, emerged as one of the primary candidate ground state of the doped Hubbard model~\cite{HF_Stripes,MasaruHF} even before its initial observation in cuprates~\cite{Abbamonte2005}. A recent unrestricted Hartree-Fock analysis~\cite{HF_Stripes2022,HF_Stripes2024} for intermediate repulsion revealed that vertical stripes are the ground state in doped Hubbard systems over a wide doping range. In the strong coupling regime, approaches such as the Gutzwiller variational method~\cite{StripeGutzwiller}, DMRG~\cite{DMRG_4leg,DMRG_6leg,HybridDMRG_Stripes,Stripes2017}, dynamical mean-field theory (DMFT)~\cite{StripeDMFT,StripeDMFT2} and slave bosons mean field theory~\cite{SlaveBosons} have all found vertical stripes around $(1/8)$ doping. These findings have been further corroborated by other numerical techniques, including infinite projected entangled pair states (iPEPS)~\cite{Stripes2017,CorboziPEPs}, auxiliary field Monte Carlo (AFQMC)~\cite{CPQMC_22,Stripes2017,Stripes2024}, variational Monte Carlo (VMC)~\cite{StripesVMC0,StripesVMC2,StripesVMC1}, and density matrix embedding theory (DMET)~\cite{Stripes2017}.

Despite broad agreement on the existence of translational symmetry breaking in the doped Hubbard model, the underlying mechanism and its relevance to cuprates remain contentious. For two leg-ladder usually no stripes are observed while SC correlations dominate~\cite{LadderVMC,AnnualRev_CompHubb}. However, a multi-method study in four-leg ladder (and higher) suggests that stripes with a charge period of $8 a_0$, without coexisting superconductivity~\cite{PRX_absSC,Stripes2017} at $(1/8)$ doping. Other studies have found charge stripes with period $5a_0$  as the most probable ground state~\cite{Huang2018} at the same doping. Furthermore, several calculations finds evidence for spatial inhomogenieties of hole-rich and  hole-poor regions in the doped Hubbard model~\cite{SandroPS,ZhangPS,DMRG_4leg,PSHanke} and spin density waves with essentially homogeneous charge density\cite{ZhangSDW}. 

However, the stability of charge density waves in the paramagnetic sector of the doped Hubbard model has yet to be thoroughly investigated. Additionally, recent study shows AFM correlations persist at higher temperatures even after stripe ordering is destroyed~\cite{StripesPRX} indicating importance of the magnetic ordering for translation symmetry breaking. Recent studies also indicate that even weak next-neighbor hopping could shift the balance in favor of uniform $d$-wave superconductivity from stripes~\cite{HongDev_Science, Stripes2024}. Thus, the observation of charge ordering independent of magnetism in cuprates is difficult to reconcile within the stripe paradigm.

We investigate density wave ordering in the paramagnetic sector of the strongly correlated Hubbard model using composite operator formalism. While the slave boson techniques extend the charge-spin separation~\cite{SlaveBosons} concept from one-dimensional models leading to electron fractionalization into emergent operators that carry portions of the electronic charge and spin, we adopt a complementary approach. Starting with the atomic Hamiltonian with no hopping ($t=0$), we generate Hubbard operators that treat electrons as a sum of quasiparticles with identical charge and spin as electrons~\cite{ovchinnikovBook}. However, such excitations do not adhere to conventional bosonic or fermionic statistics. The hopping term is then introduced through a perturbation expansion using the equation of motion approach~\cite{avella1998,avella2007,avella2011,avella2014,HarrisLange}. Our study centers on holon and doublon composite operators~\cite{hubbard1963,hubbard1964}, providing a natural framework for describing the paramagnetic sector. Such composite operator excitations have been used to examine uniform solutions in the Hubbard model, including Mott insulators~\cite{avella1998,avella2007,Haurie_2024}, d-wave superconductivity~\cite{Stanescu0,Beenen,Haurie_2024}, and antiferromagnetism~\cite{AvellaAFM}, and have recently expanded to explore orbital-selective Mott phases~\cite{avella2007two,EmileOSMP} and non-Fermi liquids~\cite{HassanPRB}. Furthermore, recent ultra-cold atom simulations of the Fermi-Hubbard model highlight significant role of doublons and holes in the doped Hubbard systems~\cite{HubbCold,DoublonHolePRL,HD_Trivedi24}.

To explore translational symmetry breaking in the doped strongly correlated Hubbard model, we develop a fully self-consistent real-space approach using the composite operator method (COM). For unidirectional charge density waves, we assume translation invariance along one direction to approach the thermodynamic limit. Finally, the full real-space treatment allows us to examine the impact of disorder on density wave solutions. Key physical quantities, such as the local density of states, provide valuable insights into the resulting solutions and offer opportunities for experimental comparison. The next section details the model and real-space composite operator formalism.

\section{Model and Method}
We work with the Hubbard Hamiltonian, which is given by
\begin{align}
&\mathcal{H}=-\sum\limits_{\langle ij\rangle,\sigma}\big(t_{ij}c^\dagger_{i\sigma}c_{j\sigma}+h.c.\big)+U\sum\limits_i \hat{n}_{i\uparrow} \hat{n}_{i\downarrow}-\mu\sum_{i,\sigma} \hat{n}_{i\sigma}.&
\end{align}
Here, $c_{i\sigma}$ $(c^\dagger_{i\sigma})$ annihilates (creates) an electron at site $i$ with spin $\sigma$, where $\sigma=\uparrow,\downarrow$ for spin-$1/2$ electrons. The first term represents hopping between neighboring sites $i$ and $j$, typically $t_{ij}=t$ if $i$ and $j$ are nearest neighbors; otherwise, zero. We set $t=1$ in this work and all our energy scales are in the units of $t$. The second term accounts for on-site repulsion between electrons, characterized by the strength $U>0$. The number operator is defined as $\hat{n}_{i\sigma}= c^\dagger_{i\sigma} c_{i\sigma}$, and the local electron density is the expectation value of the number operator, given by $n_{i\sigma}=\langle \hat{n}_{i\sigma} \rangle$. The last term represents the chemical potential $\mu$, which fixes the average electron density of the system as $n=(1/N)\sum_{i,\sigma} n_{i\sigma}$, where $N$ is the total number of lattice sites. We also define the doping of the system as $\delta=(1-n)$. We conduct our study on a two-dimensional (2D) square lattice at low temperature $T\rightarrow0$, where the inverse temperature is denoted by $\beta=1/T$, with the Boltzmann constant set to unity. Our goal is to extract the single particle electronic Greens function $\mathcal{G}(i,j,\omega)$ (and hence other physical quantities) in the strong correlation regime, i.e., $U\gg t$. We employ the equation of motion method for composite operators introduced in the next subsection. 

\subsection{Composite operator -- equation of motion}
In strong correlations, holons ($\xi$) and doublons ($\eta$) describe the quasi-fermionic excitations from the singly occupied ground state~\cite{ovchinnikovBook,hubbard1963}. 
\begin{align}
    &\xi_{i\sigma}=c_{i\sigma} (1-n_{i\overline{\sigma}}) \\ 
    &\eta_{i\sigma}=c_{i\sigma}n_{i\overline{\sigma}}
\end{align}
We define our  $2N$-component basis as 
\begin{align}
\mathbf{\Psi}=\begin{pmatrix}
\xi_{1\uparrow}, 
...,
\xi_{N\uparrow},
\eta_{1\uparrow} ,
...,
\eta_{N\uparrow}
\end{pmatrix}^T&
\label{Eq:CompOP}
\end{align}
In the paramagnetic sector, one can assume the spin rotation symmetry of the composite operators. The matrix form of the imaginary time $\tau$ Green's function becomes
\begin{align}
    \mathds{G}(\tau)=- \left\langle T_\tau \left( \mathbf{\Psi}( \tau )  \mathbf{\Psi}^\dagger(0) \right)  \right\rangle,
\end{align}
where $T_\tau$ is the usual time ordering operator. The equation of motion for the composite Green's function is given by
\begin{align}
%    -\partial_\tau \mathds{G}(\tau) = &\delta(\tau)  \left\langle \{   \mathbf{\Psi}(0),\mathbf{\Psi}^\dagger(0) \}  \right\rangle -  \left\langle T_\tau \left( \mathbf{j}( \tau )  \mathbf{\Psi}^\dagger(0) \right)  \right\rangle, \label{Eq:EOM_GIj} \\
    -\partial_\tau \mathds{G}(\tau) &=\delta(\tau) \mathds{I}(0) - \mathds{M}(\tau)
    \label{Eq:EOM_GIM}
\end{align}
where we have defined the normalization matrix $\mathds{I}$ and $\mathds{M}$-matrix as
\begin{align}
    \mathds{I} &= \left\langle \{ \Psi(0),\Psi^\dagger(0) \}  \right\rangle,    \label{Eq:EOM_IMat}\\
    \mathds{M}(\tau) &=\left\langle T_\tau \left( \mathbf{j}(\tau) \mathbf{\Psi}^\dagger(0) \right) \right\rangle. \label{Eq:EOM_MMat}
\end{align}

The current $\mathbf{j}$ is given by
%\begin{align}
   $ \mathbf{j}(\tau)=-\left[ \mathcal{H} , \mathbf{\Psi}  \right]  (\tau)$
In the composite operator formalism, the currents can be approximated as
%\begin{align}
 %   \mathbf{j}(\tau)=\mathds{E} \Psi(\tau) + \mathds{K} \delta\Phi (\tau),
%\end{align}
%Here, $\mathds{E}$ represents the Energy matrix, ensuring the current remains proportional and aligned with the defined basis. However, the second term introduces new higher-order operators. To close the equations in this order, puddles of Mott solely on the current along the composite operator, as defined in Eq. (\ref{Eq:CompOP}), disregarding higher-order terms. Consequently, we obtain:
\begin{align}
\mathbf{j}(\tau)\approx\mathds{E} \Psi(\tau)
\label{Eq:Apprxj}
\end{align}
Here, $\mathds{E}$ denotes the energy matrix, ensuring the current remains proportional to and aligned with the defined basis, which holds in the atomic limit. However, the hopping term introduces higher-order operators. To close the equations, only the current along the composite operator, as defined in Eq.~(\ref{Eq:CompOP}), is considered, while other terms are neglected. \\
Using Eq.(\ref{Eq:EOM_IMat}), Eq.(\ref{Eq:EOM_MMat}) and Eq.(\ref{Eq:Apprxj}) we get the definition of the $\mathds{E}$-matrix and $\mathds{M}(0)$-matrix
\begin{align}
    \mathds{M}(0)&= \left\langle \{ \mathbf{j}(0),\Psi^\dagger(0) \}  \right\rangle     \\
    \mathds{E}(0)&= \mathds{M}(0) \mathds{I}^{-1}
\end{align}
Putting Eq.~(\ref{Eq:Apprxj}) in Eq.~(\ref{Eq:EOM_GIM}), we obtain
\begin{align}
   -\partial_\tau \mathds{G}(\tau) &= \delta(\tau)  \mathds{I} -  \mathds{E}(0) \left\langle T_\tau \left( \mathbf{\Psi}( \tau )  \mathbf{\Psi}^\dagger(0) \right)  \right\rangle
 %   \partial_\tau \mathds{G}(\tau) &= -\delta(\tau)  \mathds{I} + \mathds{E}(0)\mathds{G}(\tau)
\end{align}
Next we can Fourier transform from imaginary time $\tau$ to Matsubara frequency $\omega_n$, and after making analytic continuation ${i \omega_n \rightarrow \omega \pm i \epsilon}$, we get advanced and retarded Green's function
\begin{align}
    \mathds{G}^{R/A}(\omega)=\left[ (\omega \pm i\epsilon) \mathds{1} - \mathds{E} \right]^{-1} \mathds{I},
    \label{Eq:Grealomega}
\end{align}

where $\mathds{1}$ is the identity matrix. Now we need to calculate the $\mathds{M}$, $\mathds{I}$ and hence $\mathds{E}$ matrix to obtain the Green's function in the frequency domain for the Hubbard model. 

\subsection{Computation of E and M matrix}
The currents for the composite operators are
\begin{align}
\mathbf{j}_{i,\uparrow}&=-\mu\xi_{i,\uparrow} 
-\sum\limits_{j}t_{ij}\left(c_{j,\uparrow}-n_{i,\downarrow}c_{j,\uparrow} +S_i^-c_{j,\downarrow}-\Delta_ic_{j,\downarrow}^\dagger\right)\\
\mathbf{j}_{i+N,\uparrow}&=-(\mu-U)\eta_{i,\uparrow}\nonumber \\ 
&+\sum\limits_j t_{ij}\left(-n_{i,\downarrow}c_{j,\uparrow}+S_i^-c_{j,\downarrow}-\Delta_i c^\dagger_{j,\downarrow}\right)
\end{align}
where we have used $S_i^-=c^\dagger_{i,\downarrow} c_{i,\uparrow}$, $S_i^+=c^\dagger_{i,\uparrow}c_{i,\downarrow}$, and  $\Delta_i=c_{i,\uparrow}c_{i,\downarrow}$. For the paramagnetic basis, we can assume $\langle n_{i,\uparrow} \rangle =\langle n_{i,\downarrow} \rangle = n_i/2$.  Using these currents we can calculate the component of the hermitian $\mathds{M}$-matrix suppressing the spin-index as follows
\begin{align}
\begin{split}\label{eq:M11}
    \mathds{M}_{i,j}={}& -\delta_{ij} \left[\mu \left(1-\frac{n_i}{2}\right) +\sum\limits_l t_{il}e_{il} \right]\\
         & -t_{ij}\left(1-\frac{n_i+n_j}{2}+p_{ij}\right),
\end{split}\\
\begin{split}\label{eq:M12}
   \mathds{M}_{i,j+N}={}& \delta_{ij}\sum\limits_lt_{il}e_{il}-t_{ij}\left(\frac{n_j}{2}-p_{ij}\right),
\end{split}\\
%\begin{split}\label{eq:M21}
%  \mathds{M}_{i+N,j,\uparrow}={}&\{\mathbf{j}_{i+N,\uparrow}, \xi^\dagger_{j,\uparrow} \} = \delta_{ij}\sum\limits_lt_{il}e_{il}-t_{ij}\left(\frac{n_i}{2}-p_{ij}\right)
%\end{split}\\
\begin{split}\label{eq:M22}
    \mathds{M}_{i+N,j+N}={}& -\delta_{ij} (\mu-U)\frac{n_i}{2} -\delta_{ij}\sum\limits_l t_{il} e_{il}-t_{ij}p_{ij},
\end{split}
\end{align}
where we have introduced the following two expectation values on each bonds,
\begin{align}
    &e_{ij}=\langle\xi_{j,\downarrow}\xi^\dagger_{i,\downarrow}\rangle-\langle\eta_{j,\downarrow}\eta^\dagger_{i,\downarrow}\rangle&\\
p_{ij}=\frac{1}{2} &\left(\langle n_{i,\downarrow}n_{j,\downarrow}\rangle +\langle S_i^-S_j^+ \rangle -\langle \Delta_i\Delta_j^\dagger\rangle + h.c.  \right)  \label{eq:pij}
\end{align}
Finally we can calculate the $I$-matrix, which is diagonal in the chosen basis
\begin{align}
&\mathds{I}_{i,j}=\delta_{i,j} \left(1-\frac{n_i}{2}\right),\\
&\mathds{I}_{i+N,j+N}=\delta_{i,j} \frac{n_i}{2}.
\end{align}

\subsection{Calculating correlation function}
 The $M$-matrix and the $I$-Matrix depends on $(5N+1)$ unknown parameters ($n_i,e_{ij},p_{ij},\mu$). Note that to evaluate the unknown parameter $n_i$ and $e_{ij}$ we only need the single particle on-site and nearest neighbor inter-site correlation functions. However, $p_{ij}$ is a two-particle correlation function and is evaluated by Roth decoupling scheme~\cite{roth1969}. In principle the correlation matrix for the full composite operator basis is given by,
\begin{align}
    \mathcal{C}_{\alpha \gamma}= -\int \frac{d\omega}{4 i \pi} \left[1+\tanh\left(\frac{\beta \omega}{2}\right) \right] \left(\mathds{G}^R(\omega))-\mathds{G}^A(\omega)\right)_{\alpha \gamma}.
    \label{Eq:FullCorr}
\end{align}
The Greek indices runs from $1$ to $2N$ and hence all the spatial and composite operator correlations can be computed.  We can compute the correlation by computing the $\omega$ integral in Eq.~(\ref{Eq:FullCorr}) analytically~\cite{avella2011}. First, we write the spectral decomposition of the ${\mathds{E}=\mathds{R} \mathds{D} \mathds{R}^{-1}}$. Here $\mathds{R}$ is the matrix of right eigenvectors of $\mathds{E}$ and $\mathds{D}$ is the diagonal eigenvalue matrix. 
%The Green's function of Eq.(\ref{Eq:Grealomega}) becomes
%\begin{align}
%${\mathds{G}^{R/A}(\omega) = \mathds{R} \Big((\omega \pm i\epsilon)\mathds{1} -\mathds{D}  \Big)^{-1}\mathds{R}^{-1} \mathds{I}}.$
%\end{align}
The difference of the retarded and advanced Green function becomes delta function integral in the limit $\epsilon\rightarrow 0$ thus using Eq.~(\ref{Eq:Grealomega}) in Eq.~(\ref{Eq:FullCorr}) can be written as 
\begin{align}
    \mathcal{C}_{\alpha \gamma}=\dfrac{1}{2}\sum_{\kappa=1}^{2N}\left[1+\tanh\left(\dfrac{\beta D_{\kappa \kappa}}{2}\right)\right] R_{\alpha \kappa}S_{\kappa \gamma}
\end{align}
where we defined $\mathds{S}=\mathds{R}^{-1}\mathds{I}$. 

However, since we restrict ourselves to nearest neighbor hoppings we require only the onsite and nearest intersite correlation function. For $m$ and $n$ integers from $m=\{0,1\}=n$ the on-site correlations and the nearest neighbor $\delta$ correlations becomes
\begin{align}
    C^{0}_{m+1,n+1}(i)&= \langle \Psi_{i+mN} \Psi^\dagger_{i+nN} \rangle=\mathcal{C}_{i+mN,i+nN},
    \label{Eq:C0}\\
     C^{\delta}_{m+1,n+1}(i)&= \langle \Psi_{i+mN} \Psi^\dagger_{i+\delta+nN} \rangle =\mathcal{C}_{i+mN,i+\delta+nN}.
     \label{Eq:Cd}
\end{align}
%where $\delta\mathds{G}(\omega)=\mathds{G}^R(\omega))-{G}^A(\omega)$. 
Using the correlation function we can write the density as
\begin{align}
    n_i=2(1-C^0_{11}(i)-C^0_{22}(i)-C^0_{12}(i)-C^0_{21}(i))
    \label{Eq:ni_re}
\end{align}
And the $e_{i,i+\delta}$ parameter becomes,
\begin{align}
    e_{i,i+\delta}=C^\delta_{11}(i)-C^\delta_{22}(i).
    \label{Eq:e_re}
\end{align}
Similarly, $p_{ij}$ can be derived in terms of the on-site and inter-site correlation function using Roth decoupling scheme presented in Appendix.~[\ref{Appendix:Roth}]. The $p_{ij}$ is composed of three distinct correlation functions, as given in Eq.~(\ref{eq:pij}). The $\langle n_{i,\downarrow}n_{j,\downarrow}\rangle$ is given in Eq.~(\ref{Eq:nn}). Similarly the $\langle S_i^-S_j^+ \rangle$ and $\langle\Delta_i\Delta_j^\dagger\rangle$ are presented in Eq.~(\ref{Eq:SS}) and Eq.~(\ref{Eq:dd}) respectively.

\subsection{Self-consistency}
In the translational invariant system, we Fourier transform to momentum space to derive self-consistency conditions (See Appendix (\ref{Appendix:Uniform})). This study focuses on unidirectional density wave patterns, where the system is uniform along one direction and modulated along the other. Thus, we perform a Fourier transform along one axis (Appendix (\ref{Appendix:Unidirection})). This reduces the number of self-consistent parameters to $(5 N_x +1)$, where $N_x$ is the number of lattice points along the $x$ direction, enabling the investigation of larger systems to access the thermodynamic limit. Such uniformity along one direction provides a considerable computational advantage to searching self-consistent solutions over the parameter space for the multiple wavevectors. Furthermore, such solutions are physically relevant as multiple methods establish unidirectional vertical stripes as a strong candidate low energy state of the doped Hubbard model~\cite{Stripes2017,Stripes2024} as discussed above. A possible extension of the study is to study all charge orderings allowed by the symmetry of the square lattice.

For a given set of external parameters $(n, U, T)$, achieving convergence to a density wave solution involves choosing a trial solution modulated as follows
\begin{align}
    n_i=n+\chi \cos(q_x x_i),
\end{align}
where $q_x$ is the trial wavevector and $\chi$ is the trial amplitude of the density modulations. Other parameters can be initialized from the uniform values. We observed that stabilizing a density wave solution requires $\chi$ to be sufficiently large such that some local density approaches half-filling. The initial choice of $q_x$ generally determines the self-consistent periodicity; or the self-consistency converges to a uniform solution for a chosen $q_x$. Nevertheless, the final ordering wavevector is identified by analyzing the self-consistent profiles, as detailed in the following subsection.

We perform the self-consistency calculations as follows. We define the $E$-matrix by assuming an initial guess for the chemical potential $\mu$, local densities $\{n_i\}$ at each site, and $\{e_{ij}\}$, $\{p_{ij}\}$ on every nearest neighbor bonds. Subsequently, we numerically diagonalize the $\mathds{E}$-matrix and determine the correlation function utilizing Eq.(\ref{Eq:C0}) and Eq.(\ref{Eq:Cd}). We update the $n_i$ using Eq.(\ref{Eq:ni_re}), recompute $e_{ij}$ using Eq.(\ref{Eq:e_re}), and obtain $p_{ij}$ from Eq.~(\ref{Eq:nn}-\ref{Eq:dd}) in Appendix~(\ref{Appendix:Roth}). The average density $n$ is fixed by the chemical potential $\mu$. If the new values differ from the initial ones, we linearly mix the old and new parameters with a mixing parameter $\alpha$. Our analysis confirms that variations in $\alpha$ do not influence the self-consistent solutions, as expected. The error $\zeta$ is defined as the maximum absolute difference in the parameters between consecutive iterations. A solution is deemed self-consistent if $\zeta < 10^{-5}$ for all variables at each site. Furthermore, we have verified that the results remain unchanged even with a stricter convergence criterion of $\zeta < 10^{-10}$. Furthermore, we provide initial self-consistent parameters and final converged solutions of the order parameters for different model parameters in the Supplementary materials~\cite{SMPRB}.

\subsection{Analysis of self-consistent solutions}
We explore converged self-consistent solutions for different values of trial wavevector $q_x$ and conduct a variational analysis to determine the state with the lowest energy per site (See Appendix.~(\ref{Appendix:GSE}). Furthermore, we obtain the bond density defined as $\tau_{ij}=\langle c^\dagger_{i\sigma} c_{j\sigma} \rangle$ to track the modulation of the bond density. The d-wave and extended s-wave component for the bond density order $\tau_{ij}$.
\begin{align}
    \tau_{d(s)}(i)=\frac{1}{4}\left(\tau_{i,i+\hat{x}} + \tau_{i,i-\hat{x}} \mp\tau_{i,i+\hat{y}} \mp \tau_{i,i-\hat{y}}   \right), \label{eq:taud}
\end{align}
To determine the ordering wavevector of the site and bond density, we begin by computing the Fourier components of the variation in the ordering. The Fourier component for the local density is denoted by $\mathcal{S}(n)$  and $\tau_d$ by $\mathcal{S}(\tau_d)$. We define the primary wavevector of CDW (dBDW), $Q_n (Q_{\tau_d})$ as the strongest Fourier component of $n_i (\tau_d)$.
%\begin{align}
%   &\mathcal{S}(n)=\frac{1}{N} \sum_{i} \exp({i \mathbf{q}.\mathbf{r}_i}) (n_i-\rho).
%    \label{eq:Sn}\\
%    &\mathcal{S}(\tau_d)=\frac{1}{N} \sum_{i} \exp({i \mathbf{q}.\mathbf{r}_i}) (\tau_d(i)-\bar{\tau}_d),
 %   \label{eq:Staud}
%\end{align}
%where $\bar{\tau_d}$ is the site average d-wave bond density. Similar, definition can be applied for $\tau_s$. 

\subsection{Scope and limitations}
Here, we summarize the approximations employed to derive the self-consistent composite operator scheme. First, we seek a paramagnetic solution by restricting the basis in Eq.(\ref{Eq:CompOP})  to a single spin state. This choice is not an inherent limitation of the method but is made to focus on paramagnetic states. To study magnetic orders, one would need to retain both spin states in the basis of Eq.~(\ref{Eq:CompOP}). Similarly the basis can be modified to study the superconducting states~\cite{Stanescu0}.

Next, a critical assumption is introduced in Eq.~(\ref{Eq:Apprxj}) to close the equation of motion by confining the current to the chosen atomic basis. Before truncating the equation of motion, the computation remains exact. However, the truncation involves neglecting higher-order terms, which necessitates ensuring that these terms do not significantly affect the key physics. Here, physical insight becomes essential to select appropriate composite operators, thus minimizing the impact of neglected terms.

In this work, we chose the holon $\xi$  and doublon $\eta$ as composite operators, as they are expected to effectively capture the excitations underdoped regime when the interaction strength is larger than the bandwidth. At half-filling, spin physics becomes more significant, leading to antiferromagnetic order~\cite{PRX_Hubbard}. However, since we have made paramagnetic assumptions, such magnetic ordering cannot be accounted for in our calculations. In the doping range explored $\delta=0.07$ to $\delta=0.2$, magnetic correlations are expected to weaken, allowing holons and doublons to dominate the physics~\cite{hubbard1964,HarrisLange}, depending on the interaction strength (See Appendix~\ref{App:Comp}). We limit the application of the method to this doping range,  as the weight of the electron in the upper Hubbard band drastically reduces with increased doping. 

From this point, two distinct approaches can be taken. Incorporating higher-order terms becomes essential if the focus is on accurately finding the ground state of the Hubbard model. However, this approach can quickly become intractable due to the exponential growth of the basis. Alternatively, one can take an approximate approach to assess whether the chosen composite operators (here, holons and doublons) provide meaningful insights into phenomena such as charge density waves. While this may not fully capture the behavior of the Hubbard model, it can still shed light on the underlying physics of a given regime, which we adopt here.

Finally, studying such composite operator methods in the one-dimensional Hubbard models where exact solutions are known is tempting. However, in lower dimensions, spin-charge separation~\cite{Essler_2005} plays a significant role. Consequently, a straightforward extension using the holon-doublon basis is unlikely to capture the physics of the 1D Hubbard model, especially when higher-order terms are neglected.

\section{Results}
We present our findings for the 2D Hubbard model on a $120\times120$ square lattice with $U=8t$. Our investigation covers doping levels ranging from $\delta=0.07$ to $\delta=0.2$, with a temperature of $T=0.01t$. In addition to examining the uniform solution, we explore self-consistent unidirectional CDW ranging from $Q_n=(1/3)$ $(2\pi/a_0)$ to very long wavelength charge modulations of $Q_n=(1/60)$ $(2\pi/a_0)$ for each set of parameters. We vary the size of the system commensurate with wavelength $\lambda_n=(2 \pi/Q_n)$ of the density wave modulations.

\begin{figure}[h]
\includegraphics[width=0.49\textwidth]{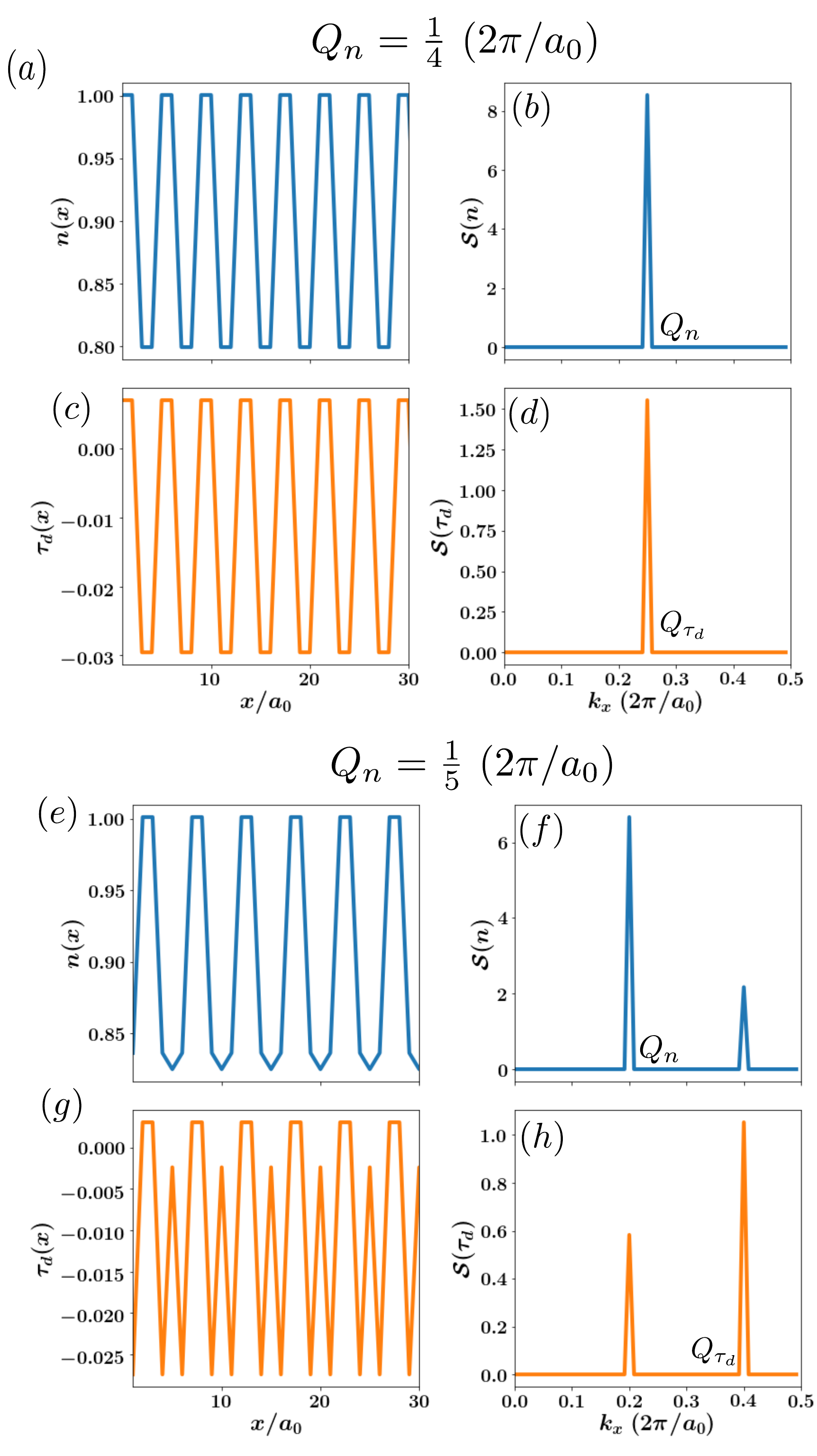}
\caption[0.5\textwidth]{Shows two representative spatial modulations and its Fourier transform for doping $\delta=0.1$. (a-d) For $Q_{n}=0.25$ and (e-h) $Q_n=0.2$. (a) Shows the modulation of the density with period $4a_0$ and (b) shows the sharp peak at ordering wavevector $Q_n=0.25$. (c) Shows the d-wave bond density modulations $\tau_d$ and (d) presents the Fourier transform of the same identifying the primary wavevector of modulation $Q_{\tau_d}=Q_n=0.25$. (e-f) Presents the density profile (Fourier components)  for $5a_0$ periodic CDW. A weak peak at $2Q_n$ is also observed. (g-h) Displays the d-wave density profile and its Fourier transform for the same solution as above with $Q_{\tau_d}=2 Q_n$.}
\label{fig:fig1}
\end{figure}

\subsection{Spatial profiles of modulated solutions}
Our analysis uncovers multiple self-consistent density wave solutions. In Fig.~(\ref{fig:fig1}), we present two such solutions for $\delta=0.1$. Fig.~(\ref{fig:fig1}a-d) illustrate the CDW solution with a periodicity of $4a_0$, while the (\ref{fig:fig1}e-h) depict the same for the $5a_0$ periodic density wave.

A common feature across all density wave solutions is the emergence of locally half-filled regions, where the local electron density reaches $n_i=1$. Since double occupancies are penalized by the repulsive interaction $U$, the density never exceeds half-filling. Such formation of Mott regions necessitates other regions to fall below the average electron density, resulting in a square-wave like CDW profile, as seen in both Fig.(\ref{fig:fig1}a) and Fig.(\ref{fig:fig1}e). The number of holes per unit cell of the CDW modulations is $n_h\sim 0.4$ for $Q_n=0.25$ and $n_h\sim 0.5$ for $Q_n=0.2$, thus indicating to a hybridization of the holes across the Mott regions. 

In Fig.(\ref{fig:fig1}b), we present the Fourier components of the CDW modulations, revealing a single peak at the anticipated ordering wavevector $k_x=0.25 (2\pi/a_0)\equiv Q_n$. This defines our primary wavevector $Q_n$ as indicated. Similarly, in Fig.(\ref{fig:fig1}f), the strongest peak appears at the expected $k_x=0.2 \equiv Q_n$, along with higher harmonics of oscillations at $k_x=0.4$.

We illustrate the spatial characteristics of the d-wave bond density $\tau_d$ in Fig.(\ref{fig:fig1}c) and Fig.(\ref{fig:fig1}f) alongside the corresponding density profiles. Remarkably, the dBDW captures the local fluctuations of the density. Due to mild density variations in the low-density regions of Fig.~(\ref{fig:fig1}), the $\tau_d$ pattern becomes distinct from the density profile. However, in our analysis, the dBDW pattern remains primarily governed by the ddensity wave structure.

In Fig.(\ref{fig:fig1}d), we observe the same dominant wavevector, such that $Q_n=Q_{\tau_d}$. However, in Fig.(\ref{fig:fig1}h), the dominant wavevector is doubled, with $Q_{\tau_d}=2 Q_n$. Similar features are observed for solutions with longer periods of modulations because they exhibit density variations in spatially hole-rich regions. The density profile deviates from the square wave as it becomes rounded in the hole-rich region for $Q_n=0.2$. The density variation, in turn, leads to oscillatory domains in $\tau_d$ in the hole-rich region in Fig.~(\ref{fig:fig1}g), leading to the doubling of periodicity of d-wave BDW.

We discuss the characteristics of the bond density in the extended s-wave, which we have not depicted here. The $\tau_s$ reduces in the spatially half-filled regions and amplifies in the hole-rich regions, mirroring the local density profile but out of phase. Consequently, the dominant ordering wavevector for sBDW is $Q_{\tau_s}=Q_n$ for all the converged solutions.

\begin{figure}[h!]
\includegraphics[width=0.475\textwidth]{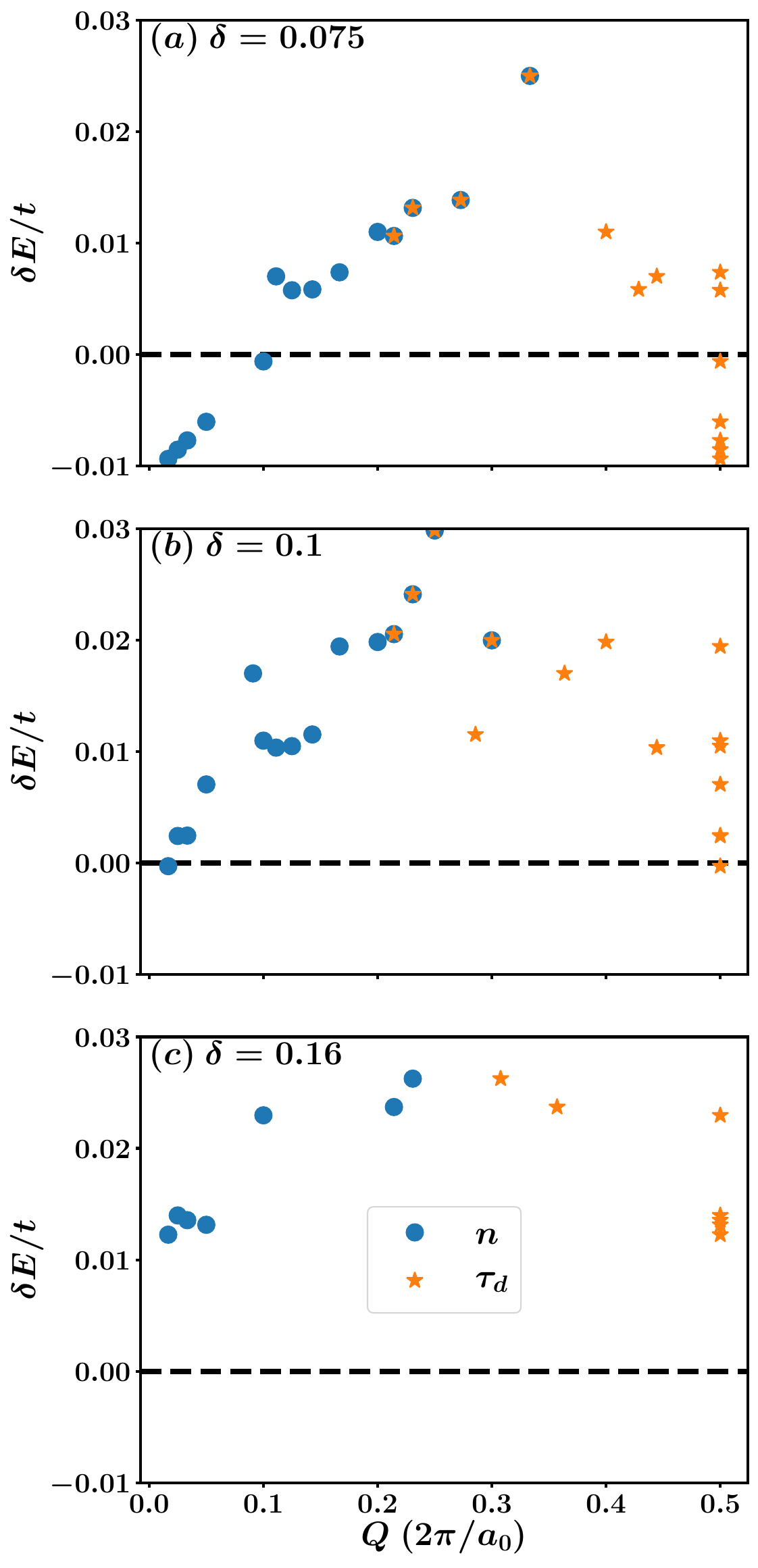}
\caption[0.5\textwidth]{Comparison of difference of energies of the modulated solution with wavevector $Q_n$ and $Q_{\tau_d}$ against the uniform solution shown in dashed lines. As doping increases the modulated solution becomes energetically unstable.}
\label{fig:fig2}
\end{figure}
\subsection{Doping dependence of energy}
In Fig.~(\ref{fig:fig2}), we compare the energy of the converged solutions for three doping values. We examine the energy difference, $\delta E = (E_{\rm uni} - E_Q)$, where $E_{\rm uni}$ represents the energy of the self-consistent uniform solution with $n_i = \rho$ for all sites, and $E_Q$ is the energy of a specific converged modulated solution. We present the energy dependence of $Q_n$ and $Q_{\tau_d}$ for particular solutions, where a single solution can possibly have a different dominant wavevector for density modulations and $\tau_d$ as shown in Fig.(\ref{fig:fig1}).

All density wave solutions lie energetically within the $\Delta E \sim 0.03 t$ range. In Fig.(\ref{fig:fig2}a) and Fig.(\ref{fig:fig2}b), we find solutions in the range $\lambda_n=4a_0$ to $3 a_0$ i.e. $Q = 0.2$ to $0.33$ where $Q_n$ and $Q_{\tau_d}$ overlap for a given solution. The same solutions have different dominant wavevectors for electron density and $\tau_d$ in the case of stripes with longer wavelength.

As doping increases, the modulated solutions become less favorable energetically, and only self-consistency with only longer periodicity is achieved. Finally, for doping $\delta=0.2$, we cannot converge to the density wave solutions. Notably, the solutions with a periodicity of $3a_0$ to $5 a_0$ become unstable at high doping, as shown in Fig.~(\ref{fig:fig2}c). 

The CDW with the largest periodicity $\lambda_n$ is energetically favorable at low dopings. Hence, our analysis reveals that strong electronic repulsion create the half-filled modulated charge density wave with a long wavelength indicating a tendency of bunching up of holes. However, for $\delta=0.07$  to $\delta=0.12$, the CDW with $3 a_0$ to $8 a_0$ period also remains energetically close to the ground state solutions. As doping increases, the uniform solutions become favored over the density wave ones. While the energetically favorable solutions are generally more relevant, the following subsections delve into the physical implications of short-wavelength density wave solutions of period $4a_0$ to $8a_0$.

\begin{figure}[ht]
\includegraphics[width=0.475\textwidth]{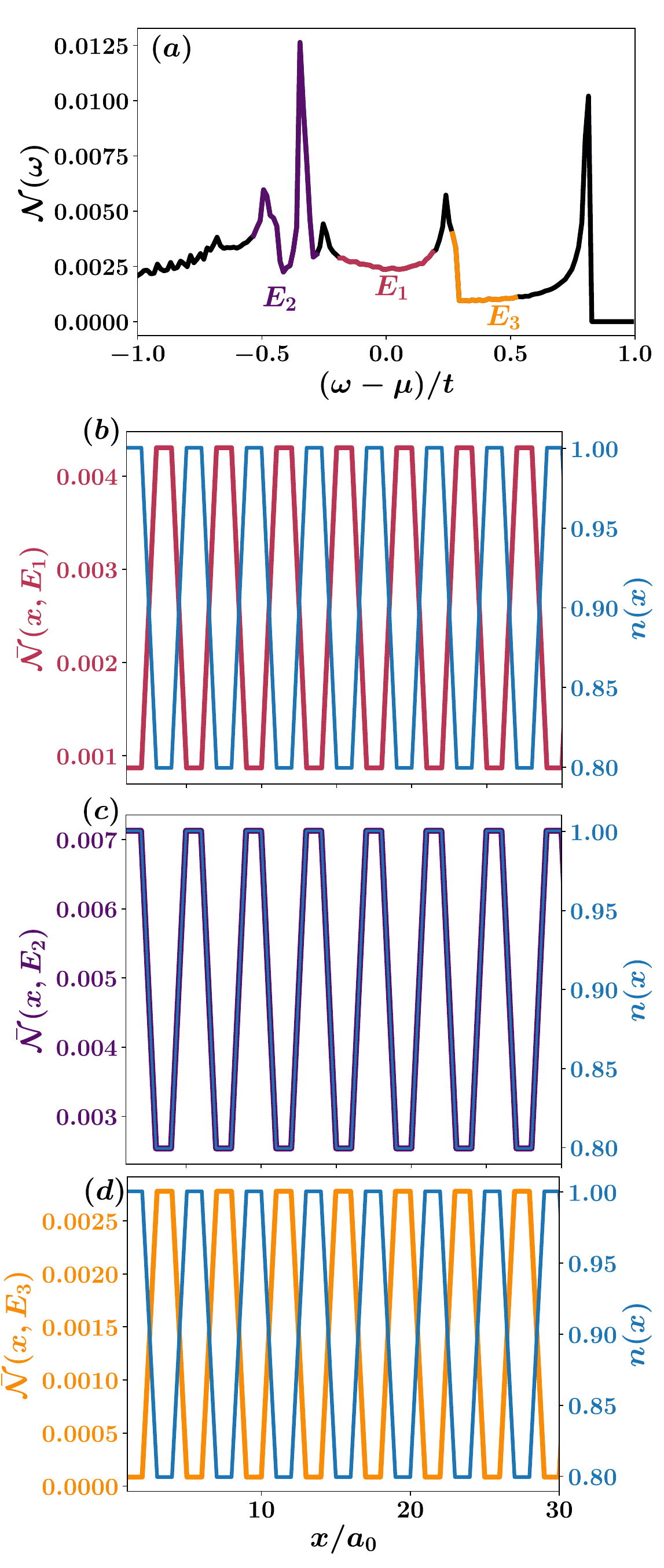}
\caption[0.5\textwidth]{(a) Shows the site-averaged density of states $\mathcal{N}(\omega)$ of the CDW solution at $\delta=0.1$ with $Q_n=0.25=Q_{\tau_d}$. Here we mark three energy ranges, $E_1$ near near the Fermi-energy. $E_2$ at the negative energy and $E_3$ at positive energy. (b) Integrated LDOS at $E_1$ (c) $E_2$ and (d) $E_3$ compared with the local electron density profile. The extraction of electrons from the Mott regions requires $\sim E_2$ energy. }
\label{fig:fig3}
\end{figure}
\subsection{Local density of states}
In Fig.~(\ref{fig:fig3}a), we display the site-averaged low-energy density of states for the $Q_n=0.25$ CDW solution at $\delta=0.1$. We only display the lower Hubbard band in the Fig.~(\ref{fig:fig3}a). At the Fermi energy no gap is observed hinting the the system remains metallic for such density wave solutions. Notably sharp excitation peaks are evident at negative energies, labeled and colored as $E_2$. We discuss below that they arise due to the local half-filled regions, i.e. it is the energy required to extract electrons from the Mott regions regions. Additionally there are no such corresponding strong peaks observed at positive energies, colored and labeled as $E_3$, as it is not possible to insert electrons without going in to the upper Hubbard band. The energy ranges are colored in Fig.~(\ref{fig:fig3}a) and next we try to explore the local density of states at these marked energy ranges. 

We present the spatially resolved local density of states (LDOS) and compare it with the local electron density profiles in Fig.~(\ref{fig:fig3}b-d). The LDOS is integrated over an energy range given by
\begin{align}
    \mathcal{\bar{N}}(E_n,x)=\int_{\omega \in E_n}  \mathcal{N}(\omega,x) d \omega
\end{align}

In Fig.~(\ref{fig:fig3}b) the low-energy excitations around the Fermi energy labeled as $E_1$ consistently emerge in the hole-rich regions. It indicates that extracting and inserting electrons in to the Mott regions is severely restricted around the $E_F$. However, up on increasing the negative bias it becomes possible to extract electrons from the half-filled regions. In Fig.(\ref{fig:fig3}c), the LDOS peaks precisely at the Mott regions, indicating that at $E_2$, electrons are extracted from the Mott regions. However, inserting electrons into the Mott regions at $E_3=-E_2$, as shown in Fig.(\ref{fig:fig3}d) is suppressed. This suppression occurs because creating double occupancies requires much larger energy excitation, $E \sim U$ leading to a pronounced asymmetry in the excitation spectrum of such strongly correlated CDW state.

\begin{figure}[h!]
\includegraphics[width=0.475\textwidth]{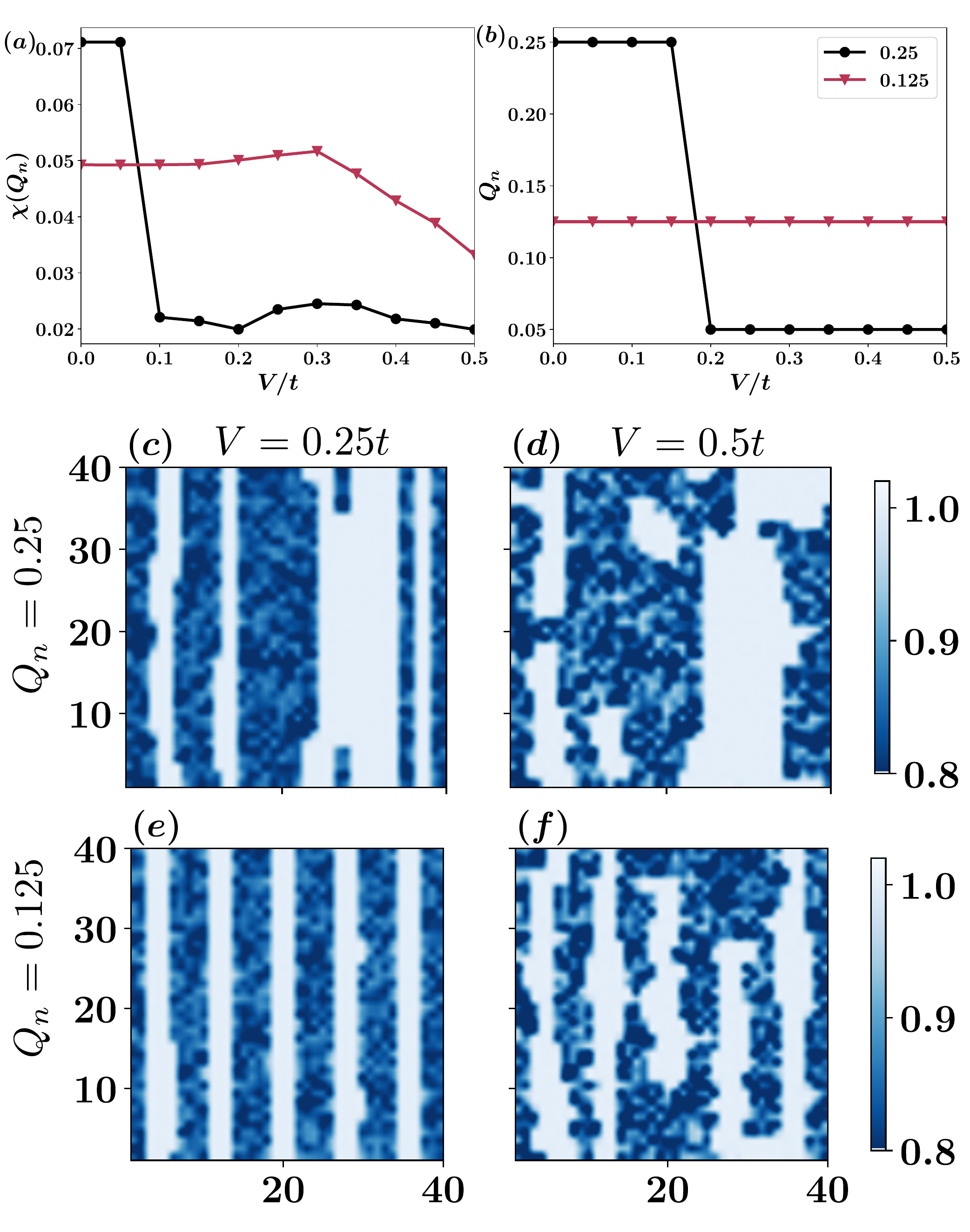}
\caption[0.5\textwidth]{(a) Displays the strength of the CDW ordering with increasing disorder strength for two different self-consistent solutions $Q_n(V=0)$ at $\delta=0.1$ in a $40 \times 40$ system. (b) Shows the ordering wavevector variation as disorder increases. The CDW with a longer wavelength demonstrates resilience to disorder. (c) Shows the spatial electron density profile with disorder $V=0.25t$ for $Q_n(0)=0.25$ (d) Displays the same for $V=0.5t$. (e) Shows local density for $V=0.3t$ for $Q_n(0)=0.125$. (f) Same for $V=0.5t$.}
\label{fig:fig4}
\end{figure}
\subsection{Effect of disorder on stripes CDW}
We then examine the effect of quenched disorder on unidirectional CDW. In cuprates, inherent inhomogeneities typically lead to short-range unidirectional density modulations (approximately $10a_0$ to $100 a_0$)~\cite{frano2020charge, VojtaCDWRev}. While quenched disorder disrupts long-range order by acting as random fields for phases with broken translational symmetry~\cite{ImryMa}, it can also pin dynamic CDW fluctuations, making them observable through local scanning tunneling probes~\cite{VojtaCDWRev}. 

We start the calculations with the self consistent density wave solution obtained in the previous section and introduce disorder $V_i$ on each lattice site. For $V=0$ and identical system size, the unidirectional and full real space calculations converge to the same solutions as expected. The disorder is randomly chosen from a uniform box distribution $V_i \in [-V/2, V/2]$, with $V$ representing the strength of the disorder. We simulate the system on a $40\times40$ lattice at $\delta=0.1$. 

We track the disorder evolution of amplitude of the CDW modulations \(\chi(Q_n)\) and the ordering wavevector. To determine these, we Fourier transform the local density and obtain $S(\mathbf{q})$. The ordering vector is determined by the position of the dominant peak of $S(\mathbf{q})$. Note that the ordering wavevector can change in the presence of disorder and generally become bidirectional $\mathbf{Q_n}$. However, the overall unidirectionality is maintained in our finite size calculation, although the $Q_n$ varies. The amplitude of the CDW modulations is given by the peak intensity, i.e., $\chi(Q_n)=S(Q_n)$. Both these quantities are evaluated by averaging over $20$ independent configurations.

Fig.~(\ref{fig:fig4}a) illustrates the evolution of the strength of CDW order $\chi(Q_n)$ as a function of disorder, for two different $Q_n$ of the clean system i.e. at $V=0$.  The strength of the CDW ordering peak diminishes upon disordering the system. However, CDWs with shorter $Q_n(V=0)$ (larger period) exhibit greater resilience to disorder.  In 2D systems, long-range CDW order is expected to be destroyed by any finite disorder, leading to inhomogeneous regions or ``puddles"~\cite{SubirCDWdis,BanerjeePRB}. The greater resilience of long wavelength CDW to the disorder indicates a tendency to form larger puddles of the coherent charge modulations only. In Fig.~(\ref{fig:fig4}b), we present the dominant wavevector of the CDW modulation. As the disorder strength increases, the ordering wavevector diminishes. However, the dominant ordering of $Q_n=0.125$ remains stable up to higher $V$. 

We depict the CDW profile for two $Q_n(0)$ values at two disorder strengths to visually compare their local effects. Interestingly, remnants of the uniform charge modulation from clean systems persist even in disordered environments, as shown in Fig~(\ref{fig:fig4}c-f). In Fig.~(Fig.(\ref{fig:fig4}c), even for weak disorder, some of the Mott regions vanish, and some half-filled regions merge, creating a larger puddle. As disorder increases, such merging occurs at several locations, and the CDW profile loses its initial coherent modulation pattern completely (Fig.(\ref{fig:fig4}d)).

For $Q_n=0.125$, merging does not occur at weak disorder due to larger hole-rich regions between the two Mott regions. However, the disorder impacts the domain wall, as evident in Fig.(\ref{fig:fig4}e). Upon increasing the disorder strength in Fig.(\ref{fig:fig4}f), the CDW local pattern becomes more inhomogeneous. Apparently, the original ordering is retained up to $V=0.5t$ which can be a finite size effect. Our results suggest that disorder destroys some Mott regions and creates larger puddles of Mott regions by merging nearby ones and hence reduces the wavevector of modulations. 

\section{Discussions}
Using the composite operator formalism for the Hubbard model, we find multiple paramagnetic unidirectional density waves within the doping range $\delta = 0.07$ to $0.2$ for $U = 8t$. We have checked that the density wave solutions becomes more energetically favorable for larger Hubbard interaction $U=12t$. In Appendix.~(\ref{App:Comp}), we show the evolution of different solutions as a function of $U$ for $\delta=0.125$ and compare it with other numerical calculations. Below $\delta = 0.07$, it becomes difficult to converge on short-period density wave solutions using our method. As doping decreases, uniform solutions become more favorable than CDW states. 

Furthermore, our analysis demonstrates a density wave pattern with local Mott regions separated by hole-rich sites. Intriguingly, stabilizing any density wave solution requires the formation of such strictly half-filled regions. Methods such as Hartree-Fock~\cite{HF_Stripes2022} and DMRG~\cite{Stripes2017,DMRG_6leg} depend on the emergence of antiferromagnetic domains to facilitate translation symmetry breaking and displays a smoother density modulations. Note that the Hubbard operator formalism starts with the atomic Hamiltonian, assuming a singly occupied state on each site due to large $U$, with hopping treated as a perturbation. Consequently, enhanced charge transfer between the Mott and hole-rich regions thus forming a smoother interface possibly requires higher-order correction terms in the composite operators. Such nearly half-filled regions are also likely to display AFM ordering at lower temperatures, similar to stripe phases. However, the translation symmetry breaking can persist independent of the spin ordering.

Our calculations reveal that unidirectional density wave orders with a large periodicity represent the most favorable state in the paramagnetic sector of the Hubbard model. This suggests a natural tendency for holes to cluster together, hinting at the possibility of phase separation into hole-rich and hole-poor regions, a concept proposed as candidate ground state for the doped Hubbard model~\cite{PSHanke,PSorella0}. Notably, uniform solutions within the composite operator formalism using Roth decomposition have exhibited negative compressibility near half-filling~\cite{Stanescu0,Beenen}, suggesting a tendency toward stripe formation or phase separation. The inhomogeneous solution in this study reveals the emergence of unidirectional density wave orders while leaving the investigation of bi-directional density wave ordering and phase separation for future work.

In contrast, recent studies indicate that density modulations with an $8a_0$ period, coexisting with antiferromagnetic spin ordering, are the most stable state~\cite{Stripes2017,Stripes2024} for $U=8t$ to $12t$ at $\delta=0.125$. Notably, longer periodicity cannot be extensively explored in those works due to limitations in system size. Numerical density wave solutions with a given periodicity are strongly biased by finite-size calculations, where commensurate lattice sizes play a crucial role. Therefore, studies with large system sizes are necessary to eliminate the boundary effect biases. Our self-consistent strong-coupling approach can handle large system sizes, though it currently overlooks higher-order effects of hopping in the current and self-energy corrections. Indeed, such corrections are expected to influence the stability of CDW with different wavevectors. 

While it is still debated whether the pure Hubbard model captures the phenomenological details of the physics of the cuprates, we demonstrate the possible, stable density wave ordering without AFM background. In cuprates, density modulations are generally observed with a much shorter periodicity of $3a_0$ to $6a_0$. Our findings indicate that quenched disorders merges the nearby Mott regions and form larger half-filled regions. Such formation of short ranged puddles of charge modulations due to impurities does not reduce the dominant periodicity observed for the clean system. We study nearest-neighbor interactions, however, it progressively kills CDWs with shorter wavelength in our formalism (See Appendix~(\ref{App:NNRep})). Past works~\cite{StripeGutzwiller} have shown that long-range Coulomb interaction can reduce the tendency of phase separation and favor density wave solutions. Additionally, the competition of phase separation of hole-rich and hole-poor regions with anti-ferromagnetism~\cite{AvellaAFM} and pairing modulation tendencies~\cite{Haurie_2024,Beenen,Stanescu0,AnuragtJ} remains to be addressed in future studies. 

\section{Acknowledgement}
The authors thank M.Vojta, S. Zhang, H. Freier, and L.Haurie for useful discussions and suggestions regarding the draft. The authors acknowledges funding from CEPIFRA (Grant No. 6704-3). The calculations are performed on the IPhT Kanta cluster. 
\appendix
\section{Details of composite operator formalism}
\subsection{Roth decoupling}
\label{Appendix:Roth}
We employ the Roth decoupling scheme~\cite{roth1969} to calculate two-point correlation functions such as density-density, spin-spin, and pair-pair correlations as expressed in Eq.~(\ref{eq:pij}), thereby determining $p_{ij}$ in terms of both on-site and intersite correlations. The formalism for the same is outlined in Ref~\cite{EmileOSMP,roth1969}. \\A bosonic operator can be written as the product of two fermionic operators. Then the idea of Roth decoupling is to consider the equation of motion of a fermionic propagator instead of a bosonic one. This is the main weakness of the decoupling as it amounts to make a choice and can lead to unphysical breaking of symmetries. But as suggested in Ref.~\cite{roth1969}, the decoupling can be symmetrized. \\
Let us illustrate the method concretely by decoupling $\langle A^p_iB^p_j\rangle$ terms with $A^p_i$ and $B^p_j$ local bilinear fermion operator and $p$ runs over the number of relevant bosonic operators that need to be decoupled. In the paramagnetic approximation and with the composite operators basis composed of holon $\xi$ and doublon $\eta$, the relevant quantities to compute are $\langle\Delta_i\Delta^\dagger_j\rangle$, $\langle n_{i\downarrow}n_{j\downarrow} \rangle$, $\langle n_{i\uparrow}n_{j\downarrow} \rangle$ and $\langle S_i^-S_j^+\rangle$. \\
In full generality, $A^{p}_i$ can be expressed as $A^{p}_i=c_{i\sigma}c^{(\dagger)}_{i\sigma'}$ with $c^{(\dagger)}_{i\sigma'}$ meaning it can be in general a creation or annihilation operator. To proceed with the decoupling, we choose to single out $c_{i\sigma}$. The decoupling will be performed using the composite $\mathbf{\Psi}$ basis, though it is important to note that this method is general and can be extended to higher-order approximations. For clarity and readability, we omit the $p$ index in the following discussion.\\
Let's consider the two-particle retarded Greens function $\mathds{D}_{\alpha jl}(\tau)$
\begin{align}
&\mathds{D}_{\alpha jl}(\tau)=-\left\langle T_\tau\left(\mathbf{\Psi}_\alpha(\tau)c_{j\sigma}^{(\dagger)}B_l\right) \right\rangle&
\end{align}
$\mathds{D}_{\alpha jl}(\tau)$ is a fermionic Green's function and the choice of Roth decoupling can be seen as taking the operators $\mathbf{\Psi}_\alpha(\tau)$ and $\left[c^{\dagger}_{j\sigma}B_l\right](0)$ at different time. We then proceed by computing the equation of motion for $\mathds{D}_{\alpha j l}(\tau)$. A key aspect of this approach is that, within the composite operator approximation, $j(\tau)=\mathds{E}\mathbf{\Psi}(\tau)$. This ensures that the equation of motion for $\mathds{D}_{\alpha j l}(\tau)$ is closed, enabling the development of a self-consistent scheme. The EoM for $\mathds{D}$ is given by
\begin{align}
-\partial_\tau \mathds{D}_{\alpha j l}(\tau)&=\delta(\tau)\left\langle \left\{ \mathbf{\Psi}_\alpha(0),c_{j\sigma}^{(\dagger)}B_l\right\} \right\rangle \nonumber\\
&-\left\langle  T_\tau \left(\mathbf{j}(\tau)c_{j\sigma}^{(\dagger)}B_l\right)\right\rangle
\end{align}
In the first term $\mathbf{f}_{\alpha jl}=\left\langle \left\{ \mathbf{\Psi}_\alpha(0),c_{j\sigma}^{(\dagger)}B_l\right\} \right\rangle$, all operators are evaluated at the same time, allowing this term to be computed explicitly. The second term is simply $\mathds{E}$ times $\mathds{D}_{\alpha jl}(\tau)$
\begin{align}
-\partial_\tau \mathds{D}_{\alpha j l}(\tau)=\delta(\tau)\mathbf{f}_{\alpha jl}+\sum\limits_\beta\mathds{E}_{\alpha\beta}\mathds{D}_{\beta jl}(\tau)
\end{align}
Next we can Fourier transform and after making analytic continuation $i\omega_n \rightarrow \omega+i\delta$ we obtain an explicit expression for $\mathds{D}_{\alpha j l}(\omega)$

\begin{align}
&\sum\limits_{\beta}(\omega \delta_{\alpha\beta}-\mathds{E}_{\alpha\beta})\mathds{D}_{\beta jl}(\omega)=\mathbf{f}_{\alpha jl}&\\[5pt]
&\mathds{D}_{\alpha jl}(\omega)=\sum\limits_\beta \left(\omega \mathds{1}-\mathds{E}\right)^{-1}_{\alpha\beta}\mathbf{f}_{\beta jl}&
\end{align}

The relationship between $\mathds{E}$ and the composite Green function $\mathds{G}$ is $\mathds{G}_{\alpha\beta}(\omega)=\sum\limits_\delta\left(\omega\mathds{1}-\mathds{E}\right)^{-1}_{\alpha\delta}\mathds{I}_{\delta\beta}$. In our approximation, $\mathds{I}$ is local $\mathds{I}_{i+nN,j+mN}\propto \delta_{ij}$ with $n,m \in \{0,1\}$ and we define $\mathds{I}_{nm}(i)\equiv \mathds{I}_{i+nN,i+mN}$, $\mathds{I}_{mn}^{-1}(i)\equiv \mathds{I}^{-1}_{i+nN,i+mN}$

\begin{align}
&\mathds{D}_{i+nN,jl}(\omega)=\sum\limits_{k,m,m'}\mathds{G}(\omega)_{i+nN,k+mN} \mathds{I}^{-1}(k)_{mm'}\mathbf{f}_{k+m'N,jl}&
\label{eq:EqD}
\end{align}

The next step is to apply the fluctuation-dissipation theorem to $\mathds{D}_{\alpha jl}(\omega)$, which provides an explicit expression for the correlation function we wanted to compute $\langle \mathbf{\Psi}_\alpha c^{(\dagger)}_{j\sigma} B_l\rangle$.
\begin{align}
\left\langle \mathbf{\Psi}_\alpha c_{j\sigma}^{(\dagger)}B_l\right\rangle=\int d\omega\left(1-n_F(\omega)\right)\dfrac{-1}{2\pi i}\big[\mathds{D}^R_{\alpha jl}(\omega)-\mathds{D}^A_{\alpha jl}(\omega)\big]
\end{align}
Integrating Eq.~\ref{eq:EqD}, this lead to the following equations 
\begin{align}
 &\langle \mathbf{\Psi}_{i+nN} c^{(\dagger)}_{j\sigma}B_l \rangle=\sum\limits_{k,m,m'}C^{k-i}_{n+1,m+1}\mathds{I}^{-1}(k)_{mm'}\mathbf{f}_{k+m'N,jl}&
\end{align}

If the three spatial indices $k,j,l$ are different $\mathbf{f}_{k+m'N,jl}=0$. This fact is reducing the sum over all the lattice sites in Eq.\ref{eq:EqD} to a sum over only two values of spatial index $k$, $k=j$ or $k=l$.

\begin{align}
\langle \mathbf{\Psi}_{i+nN} c^{(\dagger)}_{j\sigma}B_l \rangle=\sum\limits_{m,m'}&\left[C_{n+1,m+1}^{j-i}(i)\mathds{I}^{-1}(j)\mathbf{f}_{j+m'N,jl} \right. \nonumber \\
& \left. +C_{n+1,m+1}^{l-i}(i)\mathds{I}^{-1}(l)\mathbf{f}_{l+m'N,jl}\right]
\end{align}

To compute the initial correlation function $\langle A_i B_j\rangle$, we need to set $i=j$ and $l\rightarrow j$. 
\begin{align}
\langle \mathbf{\Psi}_{i+nN}c^{(\dagger)}_{i\sigma}B_j\rangle=\sum\limits_{m,m'}&\left[C_{n+1,m+1}^{0}(i)\mathds{I}^{-1}(i)\mathbf{f}_{i+m'N,ij} \right. \nonumber \\ &\left.+C_{n+1,m+1}^{j-i}(i)\mathds{I}^{-1}(j)\mathbf{f}_{j+m'N,ij}\right]
\label{eq:Final_f}
\end{align}

Equation~\ref{eq:Final_f} is the central result of the Roth decoupling method. It enables the calculation of $\langle \mathbf{\Psi}_i c^{(\dagger)}_{i\sigma}B_j\rangle$ and, subsequently, the expectation values for any $A_iB_j$ operators, where $A_i$ and $B_j$ are bilinear Fermionic operators. \\
The final step is to compute $\mathbf{f}_{i+m'N,ij}$ and $\mathbf{f}_{j+m'N,ij}$. What is found by direct computation is that $\mathbf{f}_{i+m'N,ij}$ will be proportional to some bilinear expectation value $\langle A_i^pB_j^p\rangle$ whereas $\mathbf{f}_{j+m'N,ij}$ will be equal to some nearest-neighbor correlation function. An important observation is that, within the Roth decoupling framework, all relevant two-point correlation functions associated with a specific choice of composite operator basis are inherently coupled and must be solved collectively.

Let's introduce the two particle correlation function vector restricting to the paramagnetic case for simplicity
\begin{align}
\mathbf{G}_2=\left(\langle \Delta_i\Delta_j^\dagger\rangle, \langle  S_i^-S_j^+\rangle, \langle n_{i\uparrow}n_{j\uparrow}\rangle, \langle n_{i\uparrow}n_{j\downarrow}\rangle\right)^T
\end{align}
Then the previous discussion means Eq.\ref{eq:Final_f} can be written as a linear system
\begin{align}
&\mathds{A}\mathbf{G_2}=\mathbf{B}&\\[5pt]
&\mathbf{G_2}=\mathds{A}^{-1}\mathbf{B}&
\end{align}
with $\mathds{A}$ and $\mathbf{B}$ depending only on single-particle quantities. \\

Applying this decoupling scheme we calculate the two point static correlation  $\langle n_in_{i+\delta}\rangle$, $\langle S_i^-S_{i+\delta}^+\rangle$ and $\langle \Delta_i \Delta^\dagger_{i+\delta}\rangle$
\begin{align}
\langle n_in_{i+\delta}\rangle=\dfrac{-\rho^S_{i,i+\delta}}{1-\phi^2_i}&+\left[ \dfrac{n_{i+\delta}}{2(1+\phi_i)} \right. \nonumber \\ & \left. \times \left(1-\dfrac{2(C^{0}_{11}(i)+C^{0}_{21}(i))}{2-n_i} \right) \right]
\label{Eq:nn}
\end{align}

\begin{align}
    \langle S_i^-S_{i+\delta}^+\rangle=\dfrac{-\rho^S_{i,i+\delta}}{1+\phi_i}
    \label{Eq:SS}
\end{align}

\begin{align}
\langle \Delta_i\Delta_{i+\delta}^\dagger\rangle=\dfrac{\rho^\Delta_{i,i+\delta}}{1+\phi_i}
\label{Eq:dd}
\end{align}

Here we have defined the following variables
\begin{align}
\phi_i=\dfrac{2}{n_i}\big(C^{0}_{12}(i)+C^{0}_{22}(i)\big)-\dfrac{2}{2-n_i}\big(C^{0}_{11}(i)+C^{0}_{21}(i)\big)
\end{align}
\begin{align}
\rho^\Delta_{i,i+\delta}=&\dfrac{2}{2-n_{i+\delta}}\left(C^\delta_{11}(i)+C^\delta_{21}(i)\right)\left(C^\delta_{22}(i)+C^\delta_{21}(i)\right) \nonumber \\ &+\dfrac{2}{n_{i+\delta}}\left(C^\delta_{12}(i)+C^\delta_{22}(i)\right)\left(C^\delta_{11}(i)+C^\delta_{12}(i)\right)
\end{align}
\begin{align}
\rho^S_{i,i+\delta}=&\dfrac{2}{2-n_{i+\delta}}\big(C_{11}^\delta(i)+C^\delta_{12}(i)\big)\big(C^\delta_{11}(i)+C^\delta_{21}(i)\big) \nonumber \\ 
&+\dfrac{2}{n_{i+\delta}}\big(C^\delta_{22}(i)+C^\delta_{12}(i)\big)\big(C^\delta_{22}(i)+C^\delta_{21}(i)\big)
\end{align}

The validity of the Roth decoupling method depends on the assumption that the chosen composite operator basis contains the relevant excitations. If the composite operators are very short-lived, the method will fail to produce accurate results. Conversely, if the composite operators are long-lived, the Roth decoupling approach is expected to provide reliable estimates.

\subsection{Uniform systems}
\label{Appendix:Uniform}
For translationally invariant systems we perform Fourier transform in k-space, with all $n_i=n$, $p_{ij}=p$ and $e_{ij}=e$. The M-matrix becomes,
\begin{align}
&\mathds{M}_{1,1}(\mathbf{k})=-\mu \left(1-\frac{n}{2} \right)-4te- \alpha(\mathbf{k}) \left(1-n+p\right)&\\
&\mathds{M}_{1,2}(\mathbf{k})=4te-\alpha(\mathbf{k})\left(\frac{n}{2}-p\right)=\mathds{M}_{2,1}(\mathbf{k})&\\
&\mathds{M}_{2,2}(\mathbf{k})=-(\mu-U)\frac{n}{2}-4te-\alpha(\mathbf{k})p&
\end{align}
where for square lattice with nearest neighbor hopping $\alpha(k_x,k_y)=2t\left(\cos(k_x)+\cos(k_y)\right)$. Whereas $2\times2$ I-matrix is diagonal and independent of $k$ given by
\begin{align}
\mathds{I}_D=&\begin{pmatrix}
1-\frac{n}{2}, 
\frac{n}{2}
\end{pmatrix}&
\end{align}
Following Ref.~\cite{Haurie_2024} one can arrive at the self-consistency equations. The uniform solutions remains a solution even in real space calculations as expected.

\subsection{Unidirectional translation symmetry}
\label{Appendix:Unidirection}
To study unidirectional density wave states, one can assume system has translation invaraince along say $y$ direction such that $k_y$ is a good quantum number. In this case, all quantities depends on $k_y$ and on the position along the $i=[1,N_x]$.   We can partially Fourier transform the $\mathds{M}$-matrix of size $2N_x\times2N_x$ for each $k_y$ 
\begin{align}
\mathds{M}_{i,j}(k_y)&=\delta_{ij} \left[-\mu\left(1-\frac{n_{i}}{2} \right)-2t e^y_i-\sum\limits_{l}t_{il}e^x_{il} \right] \nonumber \\
&-\alpha(k_y)\delta_{ij}\  \left(1-n_i+p^y_{i}\right) \nonumber \\ &-t_{ij}\left(1-\frac{n_i+n_j}{2}+p^x_{ij}\right),\\
\mathds{M}_{i,j+N}(k_y)&=2t\delta_{ij}e^y_{i}+\sum\limits_{l}t_{il}e^x_{il} \nonumber \\ &-t_{ij}\left(\frac{n_j}{2} -p^x_{ij}\right)-\alpha(k_y)\delta_{ij}\left(\frac{n_i}{2}-p^y_{i}\rangle\right),\\
%\mathds{M}_{i+N,j}(k_y)&=2t\delta_{ij}e^y_{i}+\sum\limits_{l}t_{il}e^x_{il}-t_{ij}\left(\frac{n_i}{2} -p^x_{ij}\right) \nonumber \\
%&-\alpha(k_y)\delta_{ij}\left(\frac{n_i}{2}-p^y_{i_x}\rangle\right),\\
\mathds{M}_{i+N,j+N}(k_y)&=-\left(\mu-U\right)\delta_{ij}\frac{n_i}{2}-\delta_{ij}\sum\limits_{l}t_{il}e_{il}^x\nonumber \\ &-2t\delta_{ij}e_{i}^y-t_{ij}p^x_{ij}-\alpha(k_y)\delta_{ij}p_{i}^y,
\end{align}
where $\alpha(k_y) = 2t \cos(k_y)$. The rest of the formalism can be derieved from Ref.~\cite{EmileOSMP,Haurie_2024} and closely following the real-space description in the main text. 
\subsection{Nearest neighbor repulsion}
\label{App:NNRep}
To simulate some non-local interactions we add another interacting term to the Hubbard Hamiltonian which is nearest neighbor repulsion between electrons, with the strength $W>0$. The Hamiltonian thus becomes,
\begin{align}
\mathcal{H}=-\sum\limits_{\langle ij\rangle,\sigma}&\left(t_{ij}c^\dagger_{i\sigma}c_{j\sigma}+h.c.\right)+U\sum\limits_i \hat{n}_{i\uparrow}\hat{n}_{i\downarrow} \nonumber \\ &+W\sum\limits_{\langle ij\rangle,\sigma\sigma'}\hat{n}_{i\sigma}\hat{n}_{j\sigma'}-\mu\sum\limits_{i,\sigma}\hat{n}_{i\sigma}.
\end{align}
We assume the following strong correlation regime $U\gg t$ and $U\gg W$. In this regime we can use the same $2N$-component paramagnetic basis ignoring higher order composite operators. The $W$-term will then introduce corrections to the $M$-matrix, denoted as $\delta_W\mathds{M}$.
\begin{align}
\delta\mathds{M}_{i,j}=&\delta_{ij}\sum\limits_lW_{il}\left(n_l -\langle n_{i,\downarrow}n_{l,\downarrow}\rangle-\langle n_{i,\downarrow}n_{l,\uparrow}\rangle \right)& \nonumber\\
&-W_{ij}\left(-\langle c_{i,\uparrow}c^\dagger_{j,\uparrow} \rangle + \langle c_{i,\uparrow}\eta^\dagger_{j,\uparrow} \rangle+ \right. \nonumber \\ &\left. \langle \eta_{i,\uparrow}c^\dagger_{j,\uparrow} \rangle- \langle \eta_{i,\uparrow}\eta^\dagger_{j,\uparrow}\rangle\right)&\\
\delta\mathds{M}_{i,j+N}=&W_{ij}\left(\langle c_{i,\uparrow}\eta^\dagger_{j,\uparrow}\rangle-\langle \eta_{i,\uparrow}\eta^\dagger_{j,\uparrow}\rangle\right)&\\
%&\delta_W\mathds{M}_{i+N,j,\uparrow}=W_{ij}\left(\langle\eta_{i,\uparrow}c^\dagger_{j,\uparrow}\rangle-\langle\eta_{i,\uparrow}\eta^\dagger_{j,\uparrow}\rangle\right)&\\[5pt]
\delta\mathds{M}_{i+N,j+N}=&\delta_{ij}\sum\limits_lW_{il}\left(\langle n_{i,\downarrow}n_s{j,\downarrow}\rangle+\langle n_{i,\downarrow}n_{j,\uparrow}\rangle\right)&\nonumber \\ 
&+W_{ij}\langle \eta_{i,\uparrow}\eta^\dagger_{j,\uparrow}\rangle&
\end{align}
Here, $W_{ij}$ is defined as $W_{ij}=W\delta_{\langle ij\rangle}$, meaning it is non-zero only for neighboring sites. Following the formalism presented in the main text we can generate density wave order for non-zero nearest neighbor repulsion.
\begin{figure}[h]
\includegraphics[width=0.475\textwidth]{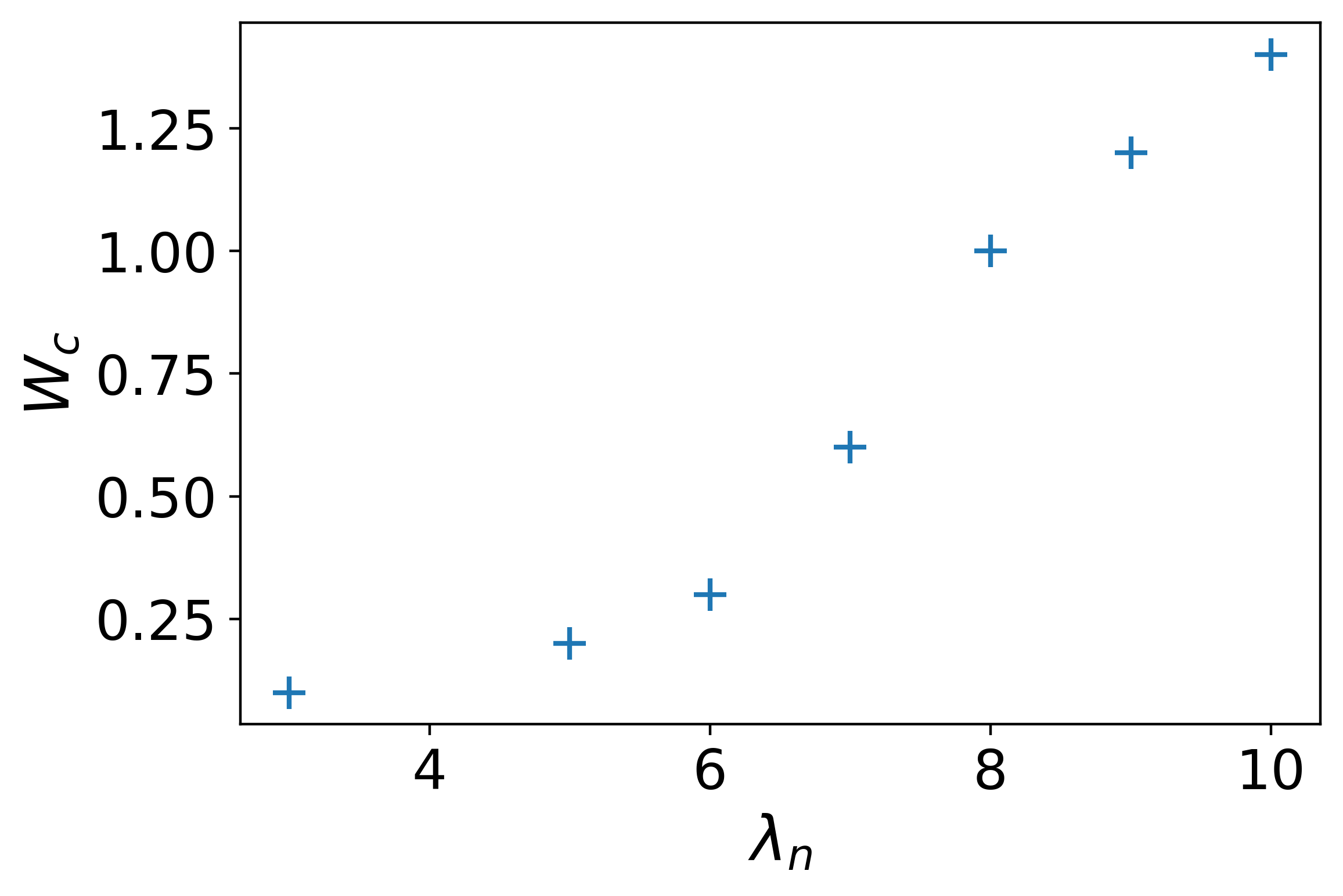}
\caption{The critical nearest neighbor repulsion for which a density wave solution of periodicity $\lambda_n$ gets killed for $U=8t$, $\delta=0.1$. The $W_c$ increases for larger wavelength $W$, indicating that shorter wavelengths cannot be favorable with just nearest neighbor repulsion.}
\label{FigApp0}
\end{figure}

Starting from the CDW solutions at $W=0$ presented in the main text, we gradually increase the nearest-neighbor repulsion and observe that the density wave solutions become unstable as this repulsion grows. In Fig.~(\ref{FigApp0}), we plot the critical value $W_c$, at which the CDW  transitions to a uniform state as a function of the density wave wavelength $\lambda_n$. The increase in $W_c$ indicates that nearest-neighbor repulsion cannot stabilize shorter-wavelength solutions by eliminating longer-wavelength ones. Additionally, we track the energy per site, showing that solutions with longer periodicity consistently have lower energy than those with shorter wavelengths.

\section{Physical quantities within composite operator method}
\subsection{Energy per site}
\label{Appendix:GSE}
The expectation value of the Hamiltonian gives the energy per site. 
\begin{align}
        E  &= -t\sum\limits_{\langle ij\rangle,\sigma} \langle  c^\dagger_{i\sigma}c_{j\sigma}+h.c.\rangle +U\sum\limits_i \langle \hat{n}_{i\uparrow} \hat{n}_{i\downarrow}\rangle 
\end{align}
The first term is the kinetic energy, and the second term is the potential energy. The kinetic energy is rewritten as
\begin{align}
    \mathcal{K}=t\sum\limits_{\langle ij\rangle,\sigma} \langle c_{j\sigma} c^\dagger_{i\sigma}  \rangle+ \langle c_{i\sigma} c^\dagger_{j\sigma} \rangle
\end{align}
Replacing $c=\xi+\eta$ and then rewriting everything in terms of correlation functions, at $T\rightarrow0$, we get,

\begin{align}
    \mathcal{K}_{i,\delta}=2 t\sum\limits_{m,n=1}^2 C^\delta_{mn}(i),
\end{align}
So total Kinetic energy becomes $\mathcal{K}=\sum_{i,\delta}  \mathcal{K}_{i,\delta}$.

Next, the potential energy is given by $\mathcal{U}=U D$, where $D$ is the double occupancy of the system
\begin{align}
    \mathcal{U}&=U\sum_i \langle c^\dagger_{i \uparrow} c_{i \uparrow} c^\dagger_{i \downarrow} c_{i \downarrow}  \rangle \nonumber \\ 
    &=U \sum_i \frac{1}{2}\left(n_i-2 C^0_{22}(i)-C^{0}_{12}(i)-C^{0}_{21}(i) \right).
\end{align}
The total energy per site is given by $E=(\mathcal{K}+\mathcal{U})/N$. 

Similarly the bond density, $\tau_{i,i+\delta}=\langle c^\dagger_{i\sigma} c_{i+\delta,\sigma}\rangle$ can be evaluated from
\begin{align}
    \tau_{i,i+\delta}=-\sum\limits_{m,n=1}^2 C^\delta_{mn}(i)
\end{align}
%The same expression can be derived from the Roth decoupling of the $\langle  \hat{n}_{i\uparrow} \hat{n}_{i\downarrow}\rangle$ term.
\subsection{Local density of states}
The electronic Green's function can be calculated from the composite Green's function  $\mathds{G}(\omega)$, by using $c=\eta+\xi$. Therefore,
\begin{align}
    \mathcal{G}(i,j,\omega)=\mathds{G}(\omega)_{i,j}&+\mathds{G}(\omega)_{i+N,j} \nonumber \\
    &+\mathds{G}(\omega)_{i,j+N}+\mathds{G}(\omega)_{i+N,j+N}
\end{align}
From this the local density of states at site $i$ can be evaluated by using,
\begin{align}
    \mathcal{N}(r_i,\omega)=-\frac{1}{\pi} \text{Im } \mathcal{G}(i,i,\omega)
\end{align}
\begin{figure}
\includegraphics[width=0.475\textwidth]{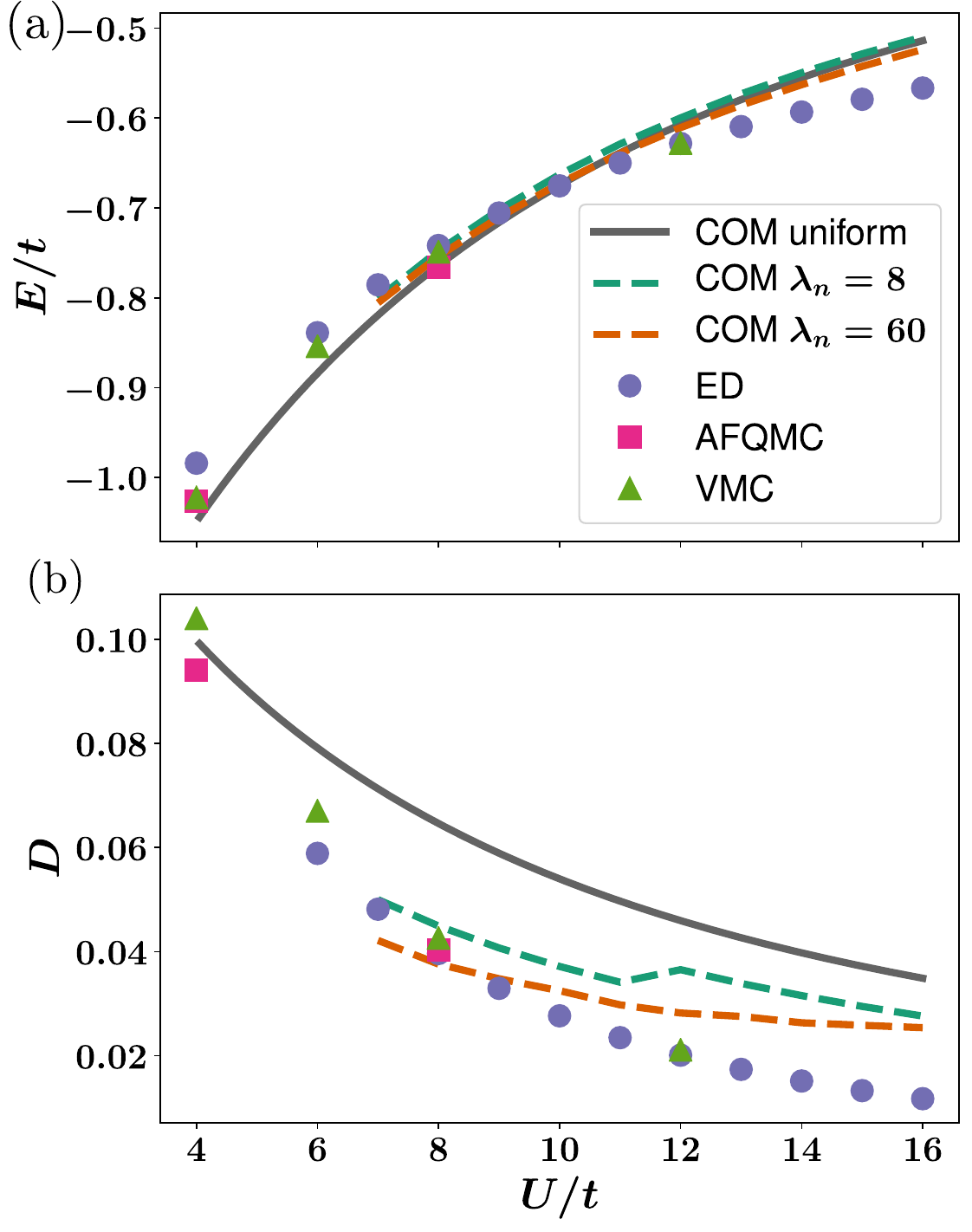}
\caption[0.5\textwidth]{(a) compares the energy per site for different solutions of the composite operator method using Roth decomposition as a function of $U$ at $n=0.875$. The energy per site at $T=0$ is shown for the uniform solution COM, along with stripes of period $\lambda_n=8$ and $\lambda_n=60$. These results are compared with those from other methods in the literature. We performed exact diagonalization (ED) on a $4\times4$ Hubbard cluster using Ref.~\cite{PRX_Hubbard} provides results from the AFQMC and VMC methods extrapolated to the thermodynamic limit from finite-size calculations. (b) compares the double occupancies from different methods.}
\label{fig:figApp1}
\end{figure}

\section{Comparison of solutions with doped Hubbard model at $n=0.875$}
\label{App:Comp}
In this appendix, we compare the Hubbard model solutions from the COM formalism presented in this work with those obtained by different methods in the literature. Since no exact method exists for treating the doped Hubbard model with large $U$ in the thermodynamic limit, these comparisons are crucial for benchmarking new solutions and improving the method. First motivated by Ref.~(\cite{PRX_Hubbard}) we compare the energy per site $E$ and the double occupancy $D$ of the Hubbard model. We choose these two quantities for a given parameter they are easily calculable for multiple methods.

\begin{table*}[ht]
\begin{tabular}{|c|c|c|c|c|c|}
\hline
$\lambda_n$ & \begin{tabular}[c]{@{}c@{}}COM\\ (This work)\end{tabular} & \begin{tabular}[c]{@{}c@{}}DMRG\\ Ref.~\cite{Stripes2017}\end{tabular} & \begin{tabular}[c]{@{}c@{}}AFQMC\\ Ref.~\cite{Stripes2017,PRX_Hubbard}\end{tabular} & \begin{tabular}[c]{@{}c@{}}iPEPs\\ Ref.~\cite{Stripes2017}\end{tabular} & \begin{tabular}[c]{@{}c@{}}DMET\\ Ref.~\cite{Stripes2017}\end{tabular} \\ \hline
1 (Uniform) & $-0.76388$                                                & ...                                                 & $-0.766 \pm 0.001$                                   & ...                                                  & $-0.7620 \pm 0.00001$                                \\ \hline
4           & ...                                                       & ...                                                 & ...                                                  & ...                                                  & $-0.7614\pm0.00005$                                  \\ \hline
5           & $-0.73384$                                                & $-0.7615 \pm 0.0004$                                & ...                                                  & $-0.7632 \pm 0.0018$                                 & $-0.7691 \pm 0.001$                                  \\ \hline
6           & $-0.74459$                                                & ...                                                 & $-0.7653 \pm 0.0002$                                 & ...                                                  & $-0.7706 \pm 0.00007$                                \\ \hline
7           & $-0.74579$                                                & $-0.762 \pm 0.001$                                  & $-0.7653\pm0.0002$                                   & $-0.7629 \pm 0.0026$                                 & $-0.7704 \pm 0.0003$                                 \\ \hline
8           & $-0.74486$                                                & $-0.762\pm0.001$                                    & $-0.7668\pm0.0002$                                   & $-0.7673 \pm 0.0020$                                 & $-0.7706 \pm 0.00001$                                \\ \hline
9           & $-0.74259$                                                & $-0.751 \pm 0.0016$                                 & $-0.7655\pm0.0002$                                   & ...                                                  & $-0.7658 \pm 0.0008$                                 \\ \hline
10          & $-0.75065$                                                & ...                                                 & $-0.7653 \pm 0.0002$                                 & ...                                                  & ...                                                 \\ \hline
60          & $-0.75530$                                               & ...                                                 & ....                                                 & ...                                                  & ...                                                 \\ \hline
\end{tabular}
\caption[0.5\textwidth]{The table compares the best estimates of energy per site (in the units of $t=1$) obtained for density wave solutions from composite operator formalism with the stripe solutions obtained by various methods in Ref.~\cite{Stripes2017} for $U=8t$ and $n=0.875$. The different methods listed are density matrix renormalization group (DMRG), AFQMC, infinite projected entangled pair states (iPEPs), and density matrix embedding theory (DMET). The system size for COM is typically $\sim 120\times120$, with translation invariance assumed along one direction. Whereas system sizes of the other methods vary from $L \times 4$ (to $L \times 8$), with the long direction $L \sim 16$ commensurate with the modulation wavelength.}
\label{TableComp}
\end{table*}

In the present work composite operator method is restricted to paramagnetic solutions of the doped Hubbard model. Proper benchmarking ideally require incorporating higher order terms and magnetic ordering, among others. Nonetheless, we present the same for both uniform and CDW solutions within our approximation, providing a useful reference for future studies.

We present in Fig.~(\ref{fig:figApp1}a) the energy per site $E$ of different self-consistent solutions from the composite operator method as a function of $U$ for $n=0.875$, a doping which is widely studied using different methods.  We present the result of density wave with domiannat periodicity $\lambda_n=4$ and $\lambda_n=60$ as a function of $U$ when after we converge self-consistently. Notably, the density wave solutions at $\delta=0.125$, only becomes favorable at large $U>11t$. 

Additionally, we include the results from exact diagonalization (ED) of $4\times 4$ clusters. The results for the ED are obtained by using the code provided in Ref.~\cite{HPhi0,HPhi1}. We also present the data obtained from Ref.~\cite{PRX_Hubbard} for uniform solutions of auxiliary field Monte Carlo (AFQMC) simulations and Variational Monte Carlo (VMC) calculations at $T=0$ are also presented. The AFQMC and VMC results extrapolated to the thermodynamic limit from finite size calculations. (See Ref.~\cite{PRX_Hubbard} and the supplemental data of the same for more details.)

For both uniform and density wave solutions, there is good agreement in energy per site with other methods in the range of $U=7t$ to $11t$. At large $U$, deviations from other methods (particularly ED) may occur due to the freezing of electrons as spins. In the limit of $U\rightarrow\infty$, particularly near half-filling, spin Hubbard operators~\cite{ovchinnikovBook} are expected to dominate as key excitations, surpassing the importance of the doublon-holon excitations considered in this study.

In Fig.~(\ref{fig:figApp1}b), we compare double occupancy as a function of $U$. At low $U$, the uniform solution remains close to the VMC and AFQMC estimates. However, the uniform solutions severely underestimate double occupancy in the intermediate to strong $U$ regime. However, the density wave solutions gives a closer $D$ with other methods indicating the need to study the the translation symmetry broken states in intermediate $U$ regimes. Such observations indicate that the holon-doublon are dominant excitations around $U\approx 8t$.

In Table~\ref{TableComp} we compare the density wave solution of commensurate wavelength $\lambda_n$ with the stripe solutions obtained from multiple methods from Ref.~\cite{Stripes2017} for $U=8t$ and $n=0.875$. The different methods listed are density matrix renormalization group (DMRG), AFQMC, infinite projected entangled pair states (iPEPs), and density matrix embedding theory (DMET). For details of these methods, refer to Ref.~\cite{Stripes2017} and references therein.

Note that for $U=8t$ and $n=0.875$, the uniform solution for the COM remains energetically favorable compared to the broken states of translational symmetry. The energy per site of the uniform solutions from COM  remains within an energy range of $\delta E \approx 0.003 t$ of the uniform solution of other methods (See . Table~(\ref{TableComp})). However, the density wave solutions from the COM with square simulation cell remain higher by energy range $\delta E\sim 0.02 t$ than the stripes obtained in cylinder geometry from DMRG, AFQMC, iPEPs, and DMET. Due to the paramagnetic assumption imposed on the density wave solutions, antiferromagnetic regions cannot form in the half-filled regions in this work. Therefore, the effect of antiferromagnetic ordering, higher-order corrections, and simulation cell geometry on the CDW solutions from COM should be investigated carefully in the future. 
\bibliography{Ref_CDW.bib}
\end{document}